\UseRawInputEncoding

\documentclass[letterpaper,10pt]{cas-dc}

\usepackage[numbers]{natbib}

\usepackage{amssymb,mathrsfs,bm} 
\usepackage{graphicx}

\usepackage[labelformat=simple]{subcaption}

\usepackage{dcolumn}
\usepackage{pifont}

\usepackage{multirow}

\usepackage{hyperref}
\hypersetup{
    colorlinks=true,
    linkcolor=blue,
    filecolor=magenta,
    urlcolor=cyan,
    citecolor=blue,
    breaklinks=true
}

\allowdisplaybreaks

\begin{document}
\let\WriteBookmarks\relax
\def\floatpagepagefraction{1}
\def\textpagefraction{.001}

\shorttitle{Flow of a viscoelastic Oldroyd-B fluid in a deformable channel} 

\shortauthors{E.~Boyko and I.~C.~Christov}

\title[mode = title]{Non-Newtonian fluid--structure interaction: Flow of a viscoelastic Oldroyd-B fluid in a deformable channel}


\author[1,2]{Evgeniy\ Boyko}[orcid=0000-0002-9202-5154]
\cormark[1]
\fnmark[1]
\ead{eboyko@purdue.edu}

\author[1]{Ivan~C.\ Christov}[orcid=0000-0001-8531-0531]
\ead{christov@purdue.edu}

\address[1]{School of Mechanical Engineering, Purdue University, West Lafayette, Indiana 47907, USA}
\address[2]{Davidson School of Chemical Engineering, Purdue University, West Lafayette, Indiana 47907, USA}

\cortext[cor1]{Corresponding author}
\fntext[fn1]{Present address: 
Faculty of Mechanical Engineering, Technion -- Israel Institute of Technology, Haifa 3200003, Israel;  evgboyko@technion.ac.il}

\begin{abstract}
We analyze the steady non-Newtonian fluid--structure interaction between the flow of an Oldroyd-B fluid and a deformable channel. Specifically, we provide a theoretical framework for calculating the leading-order effect of the fluid's viscoelasticity on the flow rate--pressure drop relation and on the deformation of the channel's elastic wall. We first identify the characteristic scales and dimensionless parameters governing the fluid--structure interaction in slender and shallow channels. Applying the lubrication approximation for the flow and employing a perturbation expansion in powers of the Deborah number $De$, we derive a closed-form expression for the pressure as a function of the non-uniform shape of the channel in the weakly viscoelastic limit up to $\mathrm{O}(De)$. Coupling the hydrodynamic pressure to the elastic deformation, we provide the leading-order effect of the interplay between the viscoelasticity of the fluid and the compliance of the channel on the pressure and deformation fields, as well as on the flow rate--pressure drop relation. For the flow-rate-controlled regime and in the weakly viscoelastic limit, we show analytically that both the compliance of the deforming top wall and the viscoelasticity of the fluid decrease the pressure drop. Furthermore, we reveal a trade-off between the influence of compliance of the channel and the fluid's viscoelasticity on the deformation. While the channel's compliance increases the deformation, the fluid's viscoelasticity decreases it.
\end{abstract}

\begin{keywords}
Fluid--structure interaction  \sep Viscoelasticity \sep 
Oldroyd-B fluid \sep Lubrication theory \sep Pressure drop \sep Microfluidics 
\end{keywords}

\maketitle

\section{Introduction}\label{Intro}

In recent years, the fluid--structure interaction (FSI) between viscous fluids and the soft deformable configurations they flow through has received considerable attention in the scientific community due to its relevance to microfluidic, lab-on-a-chip, and soft robotics applications \cite{GEGJ06,SLHLUB09,MEG17,christov2021soft}. FSI is not limited to Newtonian fluids, and it arises in various microfluidic applications involving complex fluids, such as ones containing proteins, colloidal dispersions, nucleic acids, or polymeric solutions \cite{tanner2012numerical,RS16,DGNM16,RCDC18,nahar2019influence}. In these cases, the interplay between the compliance of the confining boundaries and the complex rheological behavior of the non-Newtonian fluids involved affects the FSI in new ways that have not been fully understood. Specifically, the rheological behavior of the fluid is featured in the deformation of the soft fluidic conduit, as well as in the relationship between the pressure drop $\Delta p$ and the volumetric flow rate $q$ \cite{christov2021soft}. 

Table \ref{T1} lists a chronological selection of previous work on the steady fluid-structure interaction between complex non-Newtonian fluids and deformable configurations. From this table, we conclude that the main focus of the previous theoretical studies to date has been on shear-dependent power-law fluids. However, beyond shear thinning, complex fluids are characterized by other rheological features such as viscoelasticity, and thus, it is of fundamental and practical importance to understand how these features affect the FSI.

\begin{table*}[width=\textwidth,cols=4,pos=h]
    \caption{Chronological selection of previous experimental, numerical, and theoretical studies on the steady fluid--structure interaction of complex, non-Newtonian fluids flowing in deformable configurations.}
    \centering
    \begin{tabular*}{\tblwidth}{@{} L@{\hskip 2ex}L@{\hskip 2ex}LL@{}}
    \toprule
    Reference &  Focus  & Geometry/Solid model  & Fluid/model\\
    \midrule
    \midrule
   
    Chakraborty et al.~\cite{chakraborty2010viscoelastic,chakraborty2015viscoelastic}  &  Numer.    &  \multirow{2}{7cm}{Two-dimensional channel with a short collapsible (elastic) segment} & \multirow{2}{4cm}{Oldroyd-B, FENE-P, and Owens}\\
    
    \\[1mm]

    Yushutin \cite{yushutin2012stability}  &  Theor.    &  Axisymmetric tube as Winkler foundation & \multirow{2}{4cm}{Power law, moderate Reynolds number}\\
    
    \\[1mm]

    Tanner et al.~\cite{tanner2012numerical}  &  \multirow{2}{1cm}{Exptl./ Numer.}    &  \multirow{2}{7cm}{Thin tube with non-axisymmetric large deformation; simulations in `static' geometry} &    \multirow{2}{4cm}{Carboxymethyl-cellulose aqueous solution/Carreau} \\
    
    \\[1mm]

    Raj \& Sen \cite{RS16}  &  \multirow{2}{1cm}{Exptl./ Theor.}    &  \multirow{2}{7cm}{Slender and shallow rectangular channel with elastic  plate-like top wall}   &    \multirow{2}{4cm}{Polyethylene oxide solution/ Newtonian model only} \\
    
    \\[1mm]
    
    Del Giudice et al.~\cite{DGNM16}  &  Exptl.    &  \multirow{2}{7cm}{Square-cross-section channel extruded from large block of PDMS}   &  Polyethylene oxide solution \\
    
    \\[1mm]
  
    Raj et al.~\cite{RCDC18}  &  Exptl.    &  \multirow{2}{7cm}{Axisymmetric tube extruded from large block of PDMS}   &  Xanthan gum solution \\
    
    \\[1mm]
  
    Poroshina \& Vedeneev~\cite{poroshina2018existence}  &  Theor.    &  \multirow{2}{7cm}{Slender and thin axisymmetric shell with axial tension}   & \multirow{2}{4cm}{Power law, high Reynolds number}\\
    
    \\[1mm]
  
     Nahar et al.~\cite{nahar2019influence}  &  Exptl.    &  \multirow{2}{7cm}{Slender and thin tube, non-axisymmetric deformations}   &  \multirow{3}{4cm}{Carboxymethyl-cellulose, polyacrylamide aqueous solutions} \\
    
    \\
    
    \\[1mm]
  
    Anand et al.~\cite{anand2019non}  &  \multirow{2}{1.1cm}{Theor./ Numer.}   &  \multirow{2}{7cm}{Slender and shallow rectangular channel with elastic  plate-like top wall}   & Power law\\
    
    \\[1mm]

    Anand \& Christov \cite{anand2021revisiting}  &  \multirow{2}{1.1cm}{Theor./ Numer.}    &  \multirow{2}{7cm}{Slender and thin axisymmetric Donnell shell with axial bending}   & Power law\\
    
    \\[1mm]

    Vedeneev \cite{vedeneev2020nonlinear}  &  Theor.    &  \multirow{2}{7cm}{Geometrically nonlinear axisymmetric shell without bending; hyperelastic material}   & \multirow{2}{4 cm}{Power law, high Reynolds number}\\
    
    \\[1mm]

    Ramos-Arzola \& Bautista \cite{ramos2021fluid} &  Theor.    &  \multirow{2}{7cm}{Slender and shallow rectangular channel with elastic  plate-like top wall}    & Simplified PTT\\
    
    \\[1mm]

    \textbf{Present work} &  Theor.    &  \multirow{2}{7cm}{Slender and shallow rectangular channel with elastic  plate-like top wall}  & Oldroyd-B\\
    
    \\
    
    \bottomrule
    \end{tabular*}
    \label{T1}
\end{table*}

Recently, Ramos-Arzola \& Bautista \cite{ramos2021fluid} studied theoretically the fluid--structure interaction between a simplified Phan-Thien--Tanner (PTT) fluid \cite{thien1977new,phan1978nonlinear} flow and a slender and shallow deformable microchannel. Using lubrication theory and linear elasticity and neglecting the solvent contribution, Ramos-Arzola \& Bautista  \cite{ramos2021fluid} derived an implicit nonlinear first-order ordinary differential equation for the flow rate--pressure relation, which depends on the compliance parameter and the product $\varepsilon_\mathrm{PTT}Wi^2$, where $\varepsilon_\mathrm{PTT}$ is the extensibility parameter of the PTT model, and $Wi$ is the Weissenberg number defined in Sec.~\ref{NDsec}. For a fixed flow rate, their results predicted a decrease in the pressure drop with increasing $\varepsilon_\mathrm{PTT}Wi^2$. However, such a reduction in the pressure drop arises due to shear-thinning effects of the PTT model, which are manifested when $\varepsilon_\mathrm{PTT}Wi^2$ increases, and is consistent with results of previous theoretical studies employing the simpler shear-thinning power-law model \cite{anand2019non}. Moreover, for $\varepsilon_\mathrm{PTT}=0$, when the PTT model  corresponds to the Oldroyd-B model, the solution of Ramos-Arzola \& Bautista \cite{ramos2021fluid} for the $q-\Delta p$ relation reduces to the Newtonian relations derived by Christov et al.~\cite{christov2018flow} and Shidhore \& Christov \cite{shidhore2018static}, which are independent of the fluid's viscoelasticity. Previous investigations of an Oldroyd-B fluid in a rigid but non-uniformly shaped channel \cite{boyko2022pressure} showed that the Oldroyd-B model's flow differs from a Newtonian one, which introduces viscoelastic corrections to the pressure drop. Thus, one should anticipate that the viscoelasticity of the complex fluid affects the fluid--structure interaction, even under the Oldroyd-B model.

To the best of the authors' knowledge, the fluid--structure interaction between a constant-shear-viscosity viscoelastic (Boger) fluid and a three-dimensional deformable channel has not been analyzed in the literature, even for ``simple'' models such as Oldroyd-B and FENE-CR in the weakly viscoelastic limit, which motivates this study. 

In this work, we provide a theoretical framework for calculating the leading-order effect of the fluid's viscoelasticity on the flow rate--pressure drop relation and on the deformation of the channel's elastic wall. Our framework captures the effect of the viscoelasticity of the fluid using the Oldroyd-B model. In Sec.~\ref{PF}, we present the problem formulation and the dimensional governing equations. We further identify the characteristic scales and dimensionless parameters governing the fluid--structure interaction and provide the reduced lubrication equations for an Oldroyd-B fluid flow in a slender and shallow channel in dimensionless form. In Sec.~\ref{LA}, we present a low-Deborah-number lubrication analysis and derive a closed-form expression for the pressure as a function of the non-uniform shape of the channel up to $\mathrm{O}(De)$. Coupling the resulting expression for the pressure to the elastic deformation, in Sec.~\ref{Coupling} we provide analytical solutions for the pressure distribution and pressure drop accounting for the leading-order effect of the fluid's viscoelasticity and the channel wall's compliance.
As a concrete example of our theoretical approach, in Sec.~\ref{Results}, we present results for an Oldroyd-B fluid in a deformable channel with a compliant top wall modeled using the Kirchhoff--Love plate-bending theory. We conclude with a discussion in Sec.~\ref{CR}.

\section{Problem formulation and governing equations}\label{PF}

\begin{figure}
    \centering
    \includegraphics[width=\linewidth]{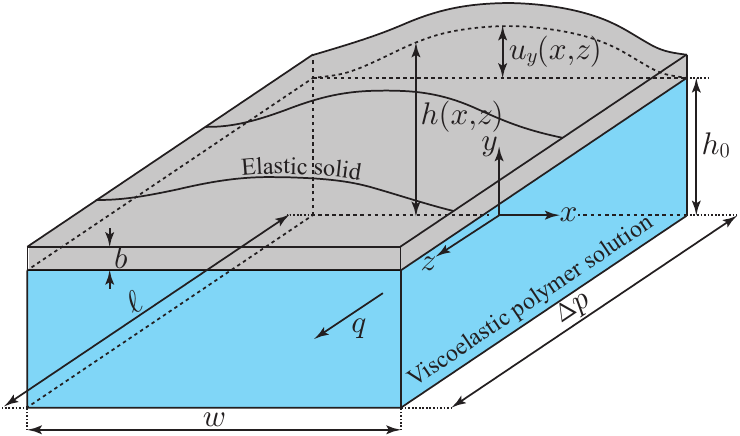}
    \caption{Schematic illustration of the problem geometry: a three-dimensional deformable channel of length $\ell$ with an initially rectangular cross-section of width $w$ and height $h_{0}$. A viscoelastic dilute polymer solution flows steadily within the channel, driven by an imposed flow rate $q$. The fluid flow stresses cause a deformation $u_{y}(x,z)$ of the fluid--solid interface, which affects the pressure drop $\Delta p$ over the axial distance $\ell$. The channel's sidewalls at $x=\pm w/2$, as well as the bottom surface at $y=0$, are assumed to be rigid. The elastic top wall has a thickness $b$.}
    \label{F1}
\end{figure}

We study the steady fluid--structure interaction between a non-Newtonian viscoelastic dilute polymer solution and a slender, shallow and deformable channel of length $\ell$, width $w$, and (deformed) height $h$, where $h \ll w \ll \ell$, as shown in Fig.~\ref{F1}. The fluid flow has a velocity field $\boldsymbol{v}=(v_{x},v_{y},v_{z})$ and pressure distribution $p$, which are induced by the imposed flow rate $q$. We seek to determine the resulting axial pressure drop $\Delta p\triangleq p(z=0)-p(z=\ell)$ for a given $q$. The channel's top wall is soft,  while its sidewalls are assumed to be rigid. As the fluid flows through the channel, the fluid stresses deform the fluid--solid interface along the channel's top wall.  We denote by $u_{y}(x,z)$ the vertical displacement of the fluid--solid interface, so that its position is given by $y=h(x,z)=h_{0}+u_{y}(x,z)$, where $h_{0}$ is the undeformed height of the channel (i.e., in the absence of the flow). We further assume that the top wall of the channel has a constant thickness $b$ and constant material properties (i.e., a Young's modulus $E_{Y}$ and a Poisson's ratio $\nu$).

We consider low-Reynolds-number flows, so that the fluid inertia is negligible compared to viscous (Newtonian and viscoelastic) stresses. In this limit, the continuity and momentum equations governing the fluid flow take the form
\begin{equation}
\boldsymbol{\nabla\cdot v}=0,\qquad\boldsymbol{\nabla\cdot\sigma}=\boldsymbol{0},\label{Continuity+Momentum}
\end{equation}
where $\boldsymbol{\sigma}$ is the stress tensor given by
\begin{equation}
\boldsymbol{\sigma}=-p\boldsymbol{I}+\underbrace{2\eta_{s}\boldsymbol{E}}_{\mathrm{Solvent}}
+\underbrace{\boldsymbol{\tau}}_{\mathrm{Polymer}}.\label{Stress tensor}
\end{equation}
The first term on the right-hand side of Eq.~(\ref{Stress tensor}) is the pressure contribution, the second term is the viscous stress contribution of the Newtonian solvent with a constant viscosity $\eta_{s}$, where $\boldsymbol{E}=(\boldsymbol{\nabla}\boldsymbol{v}+(\boldsymbol{\nabla}\boldsymbol{v)}^{\mathrm{T}})/2$ is the rate-of-strain tensor, and the last term, $\boldsymbol{\tau}$, is the polymer contribution to the stress tensor, for which a separate constitutive equation needs to be specified~\citep{bird1987dynamics1}. 

We describe the viscoelastic behavior of the complex fluid using the Oldroyd-B constitutive model~\cite{o50}, which is a well-established continuum model for viscoelastic fluids with constant shear viscosity (i.e., Boger fluids) \cite{bird1987dynamics1}. Importantly, the Oldroyd-B constitutive model can be derived from microscopic principles by modeling polymer molecules as elastic dumbbells being advected and stretched by the flow and having a linear restoring force~\cite{Intro_C_F}.
In the Oldroyd-B model, the deviatoric stress tensor is the sum of the Newtonian solvent and polymer contributions, as shown in~Eq. (\ref{Stress tensor}). At steady state, the polymer contribution $\boldsymbol{\tau}$ to the fluid's stress tensor satisfies the
upper-convected Maxwell constitutive equation
\begin{equation}
\boldsymbol{\tau}+\lambda[\boldsymbol{v}\boldsymbol{\cdot}\boldsymbol{\nabla}\boldsymbol{\tau}-(\boldsymbol{\nabla}\boldsymbol{v})^{\mathrm{T}}\boldsymbol{\cdot}\boldsymbol{\tau}-\boldsymbol{\tau}\boldsymbol{\cdot}(\boldsymbol{\nabla}\boldsymbol{v})]=2\eta_{p}\boldsymbol{E},
\label{Evolution tau_p}
\end{equation}where $\eta_{p}$ is the polymer contribution to the shear viscosity
at zero shear rate, and $\lambda$ is the longest relaxation time of the
polymers~\citep{bird1987dynamics1,Intro_C_F,alves2021numerical}. 

Using Eqs.~(\ref{Stress tensor}) and (\ref{Evolution tau_p}), the stress tensor $\boldsymbol{\sigma}$ can be also expressed as 
\begin{equation}
\boldsymbol{\sigma}=-p\boldsymbol{I}+\underbrace{2\eta_{0}\boldsymbol{E}}_{\mathrm{Newtonian}}
-\underbrace{\lambda[\boldsymbol{v}\boldsymbol{\cdot}\boldsymbol{\nabla}\boldsymbol{\tau}-(\boldsymbol{\nabla}\boldsymbol{v})^{\mathrm{T}}\boldsymbol{\cdot}\boldsymbol{\tau}-\boldsymbol{\tau}\boldsymbol{\cdot}(\boldsymbol{\nabla}\boldsymbol{v})]}_{\mathrm{Viscoelastic}},
\label{Stress tensor 2}
\end{equation}
where $\eta_{0}=\eta_{s}+\eta_{p}$ is the total zero-shear-rate viscosity of the polymer solution. 

Substituting Eq.~(\ref{Stress tensor 2}) into the second relation in Eq.~(\ref{Continuity+Momentum}) provides an alternative form of the momentum equation:  
\begin{equation}
\boldsymbol{\nabla}p=\eta_{0}\nabla^2\boldsymbol{v} -\lambda\boldsymbol{\nabla\cdot}[\boldsymbol{v}\boldsymbol{\cdot}\boldsymbol{\nabla}\boldsymbol{\tau}-(\boldsymbol{\nabla}\boldsymbol{v})^{\mathrm{T}}\boldsymbol{\cdot}\boldsymbol{\tau}-\boldsymbol{\tau}\boldsymbol{\cdot}(\boldsymbol{\nabla}\boldsymbol{v})],\label{Continuity+Momentum with S}\end{equation}which, as we show below, is convenient for assessing the viscoelastic effects on the steady flow and pressure fields of an Oldroyd-B fluid.

\subsection{Scaling analysis and non-dimensionalization}\label{NDsec}

In this work, we analyze the fluid--structure interaction of a slender and shallow deformable channel in which $h\ll w\ll \ell$. We consider a flow-rate-controlled situation, in which the characteristic axial velocity scale $v_{c}$ is set by the flow rate $q$ as $v_{c}=q/(h_{0}w)$.

We introduce dimensionless variables based on lubrication theory
\citep{tichy1996non,zhang2002surfactant,saprykin2007free,ahmed2021new,boyko2021RT,
boyko2022pressure}:%
\begin{subequations}
\begin{gather}
X=\frac{x}{w},\qquad Y=\frac{y}{h_{0}},\qquad Z=\frac{z}{\ell},\label{ND variables 1}\\
V_{x}=\frac{v_{x}}{\epsilon v_{c}},\qquad V_{y}=\frac{v_{y}}{\epsilon v_{c}},\qquad V_{z}=\frac{v_{z}}{v_{c}},\label{ND variables 2}\\
P=\frac{\epsilon^{2}\ell}{\eta_{0} v_{c}}p,\qquad H=\frac{h}{h_{0}},\qquad U=\frac{u_{y}}{u_{c}},\label{ND variables 3}\\
\mathcal{T}_{xx}=\frac{\ell}{\eta_{0} v_{c}}\tau_{xx},\quad\mathcal{T}_{yy}=\frac{\ell}{\eta_{0} v_{c}}\tau_{yy},\quad\mathcal{T}_{zz}=\frac{\epsilon^{2}\ell}{\eta_{0} v_{c}}\tau_{zz},\label{ND variables 4}\\
\mathcal{T}_{xy}=\frac{\ell}{\eta_{0}v_{c}}\tau_{xy},\quad\mathcal{T}_{xz}=\frac{\epsilon\ell}{\eta_{0}v_{c}}\tau_{xz},\quad\mathcal{T}_{yz}=\frac{\epsilon\ell}{\eta_{0}v_{c}}\tau_{yz},\label{ND variables 5}
\end{gather}\label{ND variables}\end{subequations}%
where $u_{c}$ is the characteristic scale of deformation of the top
wall and we have introduced two dimensionless parameters that quantify the slenderness and the shallowness of the channel,
\begin{equation}
\epsilon\triangleq\frac{h_{0}}{\ell}\ll1\quad\mbox{and}\quad\delta\triangleq\frac{h_{0}}{w}\ll1,\label{epsilon and delta}
\end{equation}
which are assumed to be small; the viscosity ratios,
\begin{equation}
\tilde{\beta}\triangleq\frac{\eta_{p}}{\eta_{s}+\eta_{p}}=\frac{\eta_{p}}{\eta_{0}}\quad\mbox{and}\quad\beta\triangleq1-\tilde{\beta}=\frac{\eta_{s}}{\eta_{0}},\label{beta and beta_t}
\end{equation}
and the Deborah and Weissenberg numbers, 
\begin{equation}
De\triangleq\frac{\lambda v_{c}}{\ell}=\frac{\lambda q }{w h_{0} \ell}\quad\mbox{and}\quad Wi\triangleq\frac{\lambda v_{c}}{h_{0}}=\frac{\lambda q}{w h_{0}^2}.\label{De and Wi}
\end{equation}
In Eq.~\eqref{De and Wi}, we defined the Deborah number $De$ as the ratio of the polymer relaxation time, $\lambda$, to the residence (i.e., axial advection) time in the deformable channel, $\ell/v_{c}$, or alternatively, as the product of the relaxation time scale of the fluid and the characteristic extensional rate of the flow (see \citep{tichy1996non,zhang2002surfactant,saprykin2007free,ahmed2021new}). On the other hand, we defined the Weissenberg number $Wi$ as the product of the relaxation time scale of the fluid and the characteristic shear rate of the flow, and it is related to the Deborah number through $De=\epsilon Wi$ \citep{ahmed2021new}. We note that since we have assumed that $\epsilon\ll1$, then we can have $De\ll1$ while $Wi =\mathrm{O}(1)$.

\subsection{Governing equations in dimensionless form }

Using the non-dimensionalization given in Eqs.~(\ref{ND variables})--(\ref{De and Wi}),
the governing equations (\ref{Continuity+Momentum})--(\ref{Evolution tau_p})
take the form
\begin{subequations}
\begin{equation}
\delta\frac{\partial V_{x}}{\partial X}+\frac{\partial V_{y}}{\partial Y}+\frac{\partial V_{z}}{\partial Z}=0,\label{Continuity ND}
\end{equation}
\vspace{-6mm}
\begin{multline}
\frac{\partial P}{\partial X}=\frac{\epsilon^{2}}{\delta}\left(\beta\delta^{2}\dfrac{\partial^{2}V_{x}}{\partial X^{2}}+\beta\dfrac{\partial^{2}V_{x}}{\partial Y^{2}}+\beta\epsilon^{2}\dfrac{\partial^{2}V_{x}}{\partial Z^{2}}\right.\\
 \left. +\delta\frac{\partial\mathcal{T}_{xx}}{\partial X}+\frac{\partial\mathcal{T}_{xy}}{\partial Y}+\frac{\partial\mathcal{T}_{xz}}{\partial Z}\right),\label{Momentum x ND}
\end{multline}   
\vspace{-6mm}
\begin{multline}
\frac{\partial P}{\partial Y}=\epsilon^{2}\left(\beta\delta^{2}\dfrac{\partial^{2}V_{y}}{\partial X^{2}}+\beta\dfrac{\partial^{2}V_{y}}{\partial Y^{2}}+\beta\epsilon^{2}\dfrac{\partial^{2}V_{y}}{\partial Z^{2}}\right.\\
 \left. +\delta\frac{\partial\mathcal{T}_{xy}}{\partial X}+\frac{\partial\mathcal{T}_{yy}}{\partial Y}+\frac{\partial\mathcal{T}_{yz}}{\partial z}\right),\label{Momentum y ND}
\end{multline} 
\vspace{-6mm}
\begin{multline}
\frac{\partial P}{\partial Z}=\beta\delta^{2}\dfrac{\partial^{2}V_{z}}{\partial X^{2}}+\beta\dfrac{\partial^{2}V_{z}}{\partial Y^{2}}+\beta\epsilon^{2}\dfrac{\partial^{2}V_{z}}{\partial Z^{2}} \\
 +\delta\frac{\partial\mathcal{T}_{xz}}{\partial X}+\frac{\partial\mathcal{T}_{yz}}{\partial Y}+\frac{\partial\mathcal{T}_{zz}}{\partial Z},\label{Momentum z ND}
\end{multline}   
\vspace{-6mm}
\begin{multline}
\mathcal{T}_{xx}=2\tilde{\beta}\delta\dfrac{\partial V_{x}}{\partial X} - De\left(\delta V_{x}\frac{\partial\mathcal{T}_{xx}}{\partial X}+V_{y}\frac{\partial\mathcal{T}_{xx}}{\partial Y}+V_{z}\frac{\partial\mathcal{T}_{xx}}{\partial Z}\right.\\
 \left. -2\delta\dfrac{\partial V_{x}}{\partial X}\mathcal{T}_{xx}-2\dfrac{\partial V_{x}}{\partial Y}\mathcal{T}_{xy}-2\dfrac{\partial V_{x}}{\partial Z}\mathcal{T}_{xz}\right),\label{Txx ND}
\end{multline} 
\vspace{-6mm}
\begin{multline}
\mathcal{T}_{yy}=2\tilde{\beta}\dfrac{\partial V_{y}}{\partial Y} - De\left(\delta V_{x}\frac{\partial\mathcal{T}_{yy}}{\partial X}+V_{y}\frac{\partial\mathcal{T}_{yy}}{\partial Y}+V_{z}\frac{\partial\mathcal{T}_{yy}}{\partial Z}\right.\\
 \left. -2\delta\dfrac{\partial V_{y}}{\partial X}\mathcal{T}_{xy}-2\dfrac{\partial V_{y}}{\partial Y}\mathcal{T}_{yy}-2\dfrac{\partial V_{y}}{\partial Z}\mathcal{T}_{yz}\right),\label{Tyy ND}
\end{multline} 
\vspace{-6mm}
\begin{multline}
\mathcal{T}_{zz}=2\tilde{\beta}\epsilon^{2}\dfrac{\partial V_{z}}{\partial Z} - De\left(\delta V_{x}\frac{\partial\mathcal{T}_{zz}}{\partial X}+V_{y}\frac{\partial\mathcal{T}_{zz}}{\partial Y}+V_{z}\frac{\partial\mathcal{T}_{zz}}{\partial Z}\right.\\
 \left. -2\delta\dfrac{\partial V_{z}}{\partial X}\mathcal{T}_{xz}-2\dfrac{\partial V_{z}}{\partial Y}\mathcal{T}_{yz}-2\dfrac{\partial V_{z}}{\partial Z}\mathcal{T}_{zz}\right),\label{Tzz ND}
\end{multline} 
\vspace{-6mm}
\begin{multline}
\mathcal{T}_{xy}=\tilde{\beta}\left(\dfrac{\partial V_{x}}{\partial Y}+\delta\dfrac{\partial V_{y}}{\partial X}\right) \\
- De\left(\delta V_{x}\frac{\partial\mathcal{T}_{xy}}{\partial X}+V_{y}\frac{\partial\mathcal{T}_{xy}}{\partial Y}+V_{z}\frac{\partial\mathcal{T}_{xy}}{\partial Z}+\dfrac{\partial V_{z}}{\partial Z}\mathcal{T}_{xy} \right.\\
 \left. -\delta\dfrac{\partial V_{y}}{\partial X}\mathcal{T}_{xx}-\dfrac{\partial V_{x}}{\partial Y}\mathcal{T}_{yy}-\dfrac{\partial V_{y}}{\partial Z}\mathcal{T}_{xz}-\dfrac{\partial V_{x}}{\partial Z}\mathcal{T}_{yz}\right),\label{Txy ND}
\end{multline} 
\vspace{-6mm}
\begin{multline}
\mathcal{T}_{xz}=\tilde{\beta}\left(\epsilon^{2}\dfrac{\partial V_{x}}{\partial Z}+\delta\dfrac{\partial V_{z}}{\partial X}\right) \\
- De\left(\delta V_{x}\frac{\partial\mathcal{T}_{xz}}{\partial X}+V_{y}\frac{\partial\mathcal{T}_{xz}}{\partial Y}+V_{z}\frac{\partial\mathcal{T}_{xz}}{\partial Z}+\dfrac{\partial V_{y}}{\partial Y}\mathcal{T}_{xz} \right.\\
 \left. -\delta\dfrac{\partial V_{z}}{\partial X}\mathcal{T}_{xx}-\dfrac{\partial V_{x}}{\partial Z}\mathcal{T}_{zz}-\dfrac{\partial V_{z}}{\partial Y}\mathcal{T}_{xy}-\dfrac{\partial V_{x}}{\partial Y}\mathcal{T}_{yz}\right),\label{Txz ND}
\end{multline} 
\vspace{-6mm}
\begin{multline}
\mathcal{T}_{yz}=\tilde{\beta}\left(\epsilon^{2}\dfrac{\partial V_{y}}{\partial Z}+\dfrac{\partial V_{z}}{\partial Y}\right) \\
- De\left(\delta V_{x}\frac{\partial\mathcal{T}_{yz}}{\partial X}+V_{y}\frac{\partial\mathcal{T}_{yz}}{\partial Y}+V_{z}\frac{\partial\mathcal{T}_{yz}}{\partial Z}+\delta\dfrac{\partial V_{x}}{\partial X}\mathcal{T}_{yz} \right.\\
 \left. -\dfrac{\partial V_{z}}{\partial Y}\mathcal{T}_{yy}-\dfrac{\partial V_{y}}{\partial Z}\mathcal{T}_{zz}-\delta\dfrac{\partial V_{y}}{\partial X}\mathcal{T}_{xz}-\delta\dfrac{\partial V_{z}}{\partial X}\mathcal{T}_{xy}\right).\label{Tyz ND}
\end{multline} 
 \label{ND}
\end{subequations}

\subsection{Lubrication equations for an Oldroyd-B fluid in a slender and shallow domain}

For a slender and shallow deformable channel, we assume 
the ordering $0<\epsilon\ll\delta\ll1$.
We are interested in the leading-order asymptotic behavior under this
ordering and, from Eqs.~(\ref{ND}), we obtain
\begin{subequations}
\begin{equation}
\frac{\partial V_{y}}{\partial Y}+\frac{\partial V_{z}}{\partial Z}=0,\label{Continuity LO}
\end{equation}
\begin{equation}
\frac{\partial P}{\partial Z}=\beta\dfrac{\partial^{2}V_{z}}{\partial Y^{2}}+\frac{\partial\mathcal{T}_{yz}}{\partial Y}+\frac{\partial\mathcal{T}_{zz}}{\partial Z},\quad\frac{\partial P}{\partial X}=\frac{\partial P}{\partial Y}=0,\label{Momentum LO}\end{equation}
\vspace{-6mm}
\begin{multline}
\mathcal{T}_{zz}=-De\left(V_{y}\frac{\partial\mathcal{T}_{zz}}{\partial Y}+V_{z}\frac{\partial\mathcal{T}_{zz}}{\partial Z}\right.\\
 \left. -2\dfrac{\partial V_{z}}{\partial Y}\mathcal{T}_{yz}-2\dfrac{\partial V_{z}}{\partial Z}\mathcal{T}_{zz}\right),\label{Tzz LO}
\end{multline}   
\vspace{-6mm}
\begin{multline}
\mathcal{T}_{yz}=\tilde{\beta}\dfrac{\partial V_{z}}{\partial Y}-De\left(V_{y}\frac{\partial\mathcal{T}_{yz}}{\partial Y}+V_{z}\frac{\partial\mathcal{T}_{yz}}{\partial Z}\right.\\
 \left. -\dfrac{\partial V_{z}}{\partial Y}\mathcal{T}_{yy}-\dfrac{\partial V_{y}}{\partial Z}\mathcal{T}_{zz}\right),\label{Tyz LO}
\end{multline}
\vspace{-6mm}
\begin{multline}
\mathcal{T}_{yy}=2\tilde{\beta}\dfrac{\partial V_{y}}{\partial Y}-De\left(V_{y}\frac{\partial\mathcal{T}_{yy}}{\partial Y}+V_{z}\frac{\partial\mathcal{T}_{yy}}{\partial Z}\right.\\
 \left. -2\dfrac{\partial V_{y}}{\partial Y}\mathcal{T}_{yy}-2\dfrac{\partial V_{y}}{\partial Z}\mathcal{T}_{yz}\right).\label{Tyy LO}
\end{multline}    
\label{LO Lub Eq}\end{subequations}
We note that since the leading-order lubrication equations~(\ref{LO Lub Eq}) are independent of the velocity and stress components in the transverse $x$-direction, i.e., $V_{x}$, $\mathcal{T}_{xx}$, $\mathcal{T}_{xy}$, and $\mathcal{T}_{xz}$, we do not write their corresponding evolution equations.

In fact, we have reduced the governing equations from  three dimensions (3D) to a  two-dimensional (2D) set of equations, similarly to Ahmed \& Biancofiore \cite{ahmed2021new} and Boyko \& Stone \cite{boyko2022pressure}, who analyzed flows of an Oldroyd-B fluid in non-uniform 2D rigid configurations. From Eq.~(\ref{Momentum LO}), it follows that $P=P(Z)$, i.e., the pressure is independent of $X$ and $Y$ at the leading order in $\epsilon$ and $\delta$, consistent with the classical lubrication approximation. 

\section{Low-Deborah-number lubrication analysis}\label{LA}

In the previous section, we derived the leading-order dimensionless lubrication equations (\ref{LO Lub Eq}) for an Oldroyd-B fluid, which are characterized by the two dimensionless parameters: $De$ and $\tilde{\beta}$ (or $\beta$). In this section, we consider the weakly viscoelastic limit, $De\ll1$, and derive closed-form expressions for the velocity field and the $q-\Delta p$ relation for the pressure-driven flow of an Oldroyd-B fluid through a slender and shallow deformable channel. To this end, we seek a solution of the form
\begin{equation}
\left(\begin{array}{c}
V_{z}\\
V_{y}\\
P\\
\mathcal{T}_{zz}\\
\mathcal{T}_{yy}\\
\mathcal{T}_{yz}
\end{array}\right)=\left(\begin{array}{c}
V_{z,0}\\
V_{y,0}\\
P_{0}\\
\mathcal{T}_{zz,0}\\
\mathcal{T}_{yy,0}\\
\mathcal{T}_{yz,0}
\end{array}\right)+De\left(\begin{array}{c}
V_{z,1}\\
V_{y,1}\\
P_{1}\\
\mathcal{T}_{zz,1}\\
\mathcal{T}_{yy,1}\\
\mathcal{T}_{yz,1}
\end{array}\right)+\mathrm{O}(De^{2}),\label{ND GE}
\end{equation}
and we derive analytical expressions for the velocity and stress components of interest, as well as the pressure drop, up to $\mathrm{O}(De)$. We note that, under the lubrication approximation in the low-$De$ regime considered here, it is sufficient to apply the boundary conditions,
\begin{subequations}
\begin{equation}
V_{z}(X,0,Z)=V_{z}(X,H(X,Z),Z)=0, \label{BC ND LO1}
\end{equation}
\begin{equation}
 V_{y}(X,0,Z)=V_{y}(X,H(X,Z),Z)=0,\label{BC ND LO2}
\end{equation}
\begin{equation}
\int_{-1/2}^{+1/2}\int_{0}^{H(X,Z)}V_{z}(X,Y,Z) \,\mathrm{d}Y\mathrm{d}X=1,\label{BC ND LO3}
\end{equation}
\end{subequations}
to calculate the flow and pressure fields using the asymptotic expansion given in Eq.~(\ref{ND GE}) (see \cite{black1976converging,boyko2022pressure}). In addition, we assume zero gauge pressure at the outlet of the channel and set $P(1)=0$, so that $\Delta P= P(0)$ \cite{christov2018flow,ramos2021fluid}.

We also note that the leading-order lubrication equations (\ref{LO Lub Eq}) cannot satisfy the no-slip boundary condition $V_{z}(\pm1/2,Y,Z)=0$ at the sidewalls $X=\pm1/2$. Satisfying these sidewall boundary conditions for $V_z$ requires accounting for the corrections at $\mathrm{O}(\delta)$, as discussed by Christov et al.~\cite{christov2018flow} and Boyko et al.~\cite{boyko2022flow}.

\subsection{Leading-order solution}\label{LO section}

Substituting Eq.~(\ref{ND GE}) into Eqs.~(\ref{LO Lub Eq}) and considering the
leading order in $De$, we obtain\begin{subequations}
\begin{equation}
\frac{\partial V_{y,0}}{\partial Y}+\frac{\partial V_{z,0}}{\partial Z}=0,\label{Continuity LO De}
\end{equation}
\begin{equation}
\frac{\partial P_{0}}{\partial Z}=\beta\dfrac{\partial^{2}V_{z,0}}{\partial Y^{2}}+\frac{\partial\mathcal{T}_{yz,0}}{\partial Y}+\frac{\partial\mathcal{T}_{zz,0}}{\partial Z},\label{Momentum LO De}
\end{equation}
\begin{equation}
\frac{\partial P_{0}}{\partial X}=\frac{\partial P_{0}}{\partial Y}=0,\label{Momentum X and LO De}
\end{equation}
\begin{equation}
\mathcal{T}_{zz,0}=0,\label{Tzz LO De}
\end{equation}
\begin{equation}
\mathcal{T}_{yz,0}=\tilde{\beta}\dfrac{\partial V_{z,0}}{\partial Y},\label{Tyz LO De}
\end{equation}
\begin{equation}
\mathcal{T}_{yy,0}=2\tilde{\beta}\dfrac{\partial V_{y,0}}{\partial Y},\label{Tyy LO De}
\end{equation}
\end{subequations}subject to the boundary conditions\begin{subequations}
\begin{equation}
V_{z,0}(X,0,Z)=V_{z,0}(X,H(X,Z),Z)=0, \label{BC ND LO De1}
\end{equation}
\begin{equation}
 V_{y,0}(X,0,Z)=V_{y,0}(X,H(X,Z),Z)=0,\label{BC ND LO De2}
\end{equation}
\begin{equation}
\int_{-1/2}^{+1/2}\int_{0}^{H(X,Z)}V_{z,0}(X,Y,Z) \,\mathrm{d}Y\mathrm{d}X=1.\label{BC ND LO De3}
\end{equation}
\end{subequations}
Substituting Eqs.~(\ref{Tzz LO De}) and (\ref{Tyz LO De})
into Eq.~(\ref{Momentum LO De}) yields
\begin{equation}
\frac{\mathrm{d}P_{0}}{\mathrm{d}Z}=\dfrac{\partial^{2}V_{z,0}}{\partial Y^{2}},\label{Momentum LO De e}
\end{equation}
where we have used the fact that $P_{0}=P_{0}(Z)$. As expected, at
the leading order in $De$, Eq.~(\ref{Momentum LO De}) reduces to the
dimensionless momentum equation of a Newtonian fluid with a constant
viscosity $\eta_{0}$, namely Eq.~(\ref{Momentum LO De e}).

Integrating Eq.~(\ref{Momentum LO De e}) twice with respect to $Y$ and
applying the no-slip boundary conditions Eq.~(\ref{BC ND LO De1}), we
obtain the axial velocity at the leading order:
\begin{equation}
V_{z,0}(X,Y,Z)=-\frac{1}{2}\frac{\mathrm{d}P_{0}}{\mathrm{d}Z}Y(H(X,Z)-Y).\label{Axial vel LO De}
\end{equation}
The pressure gradient $\mathrm{d}P_{0}/\mathrm{d}Z$ follows from
applying the integral constraint in Eq.~(\ref{BC ND LO De3}),
\begin{equation}
\frac{\mathrm{d}P_{0}}{\mathrm{d}Z}=-\frac{12}{\int_{-1/2}^{+1/2}H(X,Z)^{3}\mathrm{d}X}\triangleq-\frac{12}{H_{e}(Z)^{3}},\label{dP/dX LO De e}
\end{equation}
where we have defined the dimensionless effective channel height
as in \citep{wang2021reduced}:
\begin{equation}
H_{e}(Z)\triangleq\left[\int_{-1/2}^{+1/2}H(X,Z)^{3} \,\mathrm{d}X\right]^{1/3}.\label{He(Z)}
\end{equation}
For a given shape of the top wall of the channel, Eq.~(\ref{dP/dX LO De e})
provides an explicit expression for the leading-order pressure gradient. The corresponding
axial velocity distribution is then
\begin{equation}
V_{z,0}(X,Y,Z)=6\frac{Y(H(X,Z)-Y)}{H_{e}(Z)^{3}}.\label{Axial vel LO De e}
\end{equation}
Finally, the $yy$- and $yz$-components of the polymer stress tensor
at the leading order in $De$ depend on the channel shape as
\begin{subequations}
\begin{align}
   \mathcal{T}_{yy,0}&=2\tilde{\beta}\dfrac{\partial V_{y,0}}{\partial Y}=-2\tilde{\beta}\dfrac{\partial V_{z,0}}{\partial Z} \label{Tyy  LO e}\\ 
   &=-\frac{12\tilde{\beta}Y}{H_{e}(Z)^{3}}\left[\frac{3(Y-H(X,Z))H_{e}'(Z)}{H_{e}(Z)}+\frac{\partial H}{\partial Z}\right],\nonumber\\
    \mathcal{T}_{yz,0}&=\tilde{\beta}\dfrac{\partial V_{z,0}}{\partial Y}=6\tilde{\beta}\frac{H(X,Z)-2Y}{H_{e}(Z)^{3}},\label{Tyz LO e}
\end{align}\end{subequations}
where primes denote derivatives with respect to $Z$.

\subsection{First-order solution}\label{FO section}

At the next order, $\mathrm{O}(De)$, the governing equations (\ref{LO Lub Eq}) yield\begin{subequations}
\begin{equation}
\frac{\partial V_{z,1}}{\partial Z}+\frac{\partial V_{y,1}}{\partial Y}=0,\label{Continuity FO De}
\end{equation}
\begin{equation}
\frac{\partial P_{1}}{\partial Z}=\beta\dfrac{\partial^{2}V_{z,1}}{\partial Y^{2}}+\frac{\partial\mathcal{T}_{yz,1}}{\partial Y}+\frac{\partial\mathcal{T}_{zz,1}}{\partial Z},\label{Momentum FO De}
\end{equation}
\begin{equation}
\frac{\partial P_{1}}{\partial X}=\frac{\partial P_{1}}{\partial Y}=0,\label{Momentum X and Y FO De}
\end{equation}
\begin{equation}
\mathcal{T}_{zz,1}=2\dfrac{\partial V_{z,0}}{\partial Y}\mathcal{T}_{yz,0},\label{Tzz FO De}
\end{equation}
\vspace{-6mm}
\begin{multline}
\mathcal{T}_{yz,1}=\tilde{\beta}\dfrac{\partial V_{z,1}}{\partial Y}-\left(V_{y,0}\frac{\partial\mathcal{T}_{yz,0}}{\partial Y}  \right.\\
 \left. +V_{z,0}\frac{\partial\mathcal{T}_{yz,0}}{\partial Z}-\dfrac{\partial V_{z,0}}{\partial Y}\mathcal{T}_{yy,0}\right),\label{Tyz FO De}
\end{multline}
\vspace{-6mm}
\begin{multline}
\mathcal{T}_{yy,1}=2\tilde{\beta}\dfrac{\partial V_{y,1}}{\partial Y}-\left(V_{y,0}\frac{\partial\mathcal{T}_{yy,0}}{\partial Y}+V_{z,0}\frac{\partial\mathcal{T}_{yy,0}}{\partial Z}\right.\\
 \left. -2\dfrac{\partial V_{y,0}}{\partial Y}\mathcal{T}_{yy,0}-2\dfrac{\partial V_{y,0}}{\partial Z}\mathcal{T}_{yz,0}\right),\label{Tyy FO De}
\end{multline}\label{FO De}
\end{subequations}where we have used Eq.~(\ref{Tzz LO De}) to simplify
Eqs.~(\ref{Tzz FO De})--(\ref{Tyz FO De}).

These governing equations (\ref{FO De}) are supplemented by the boundary conditions\begin{subequations}
\begin{equation}
V_{z,1}(X,0,Z)=V_{z,1}(X,H(X,Z),Z)=0,\label{BC ND FO De1}
\end{equation}
\begin{equation}
 V_{y,1}(X,0,Z)=V_{y,1}(X,H(X,Z),Z)=0,\label{BC ND FO De2}
\end{equation}
\begin{equation}
\int_{-1/2}^{+1/2}\int_{0}^{H(X,Z)}V_{z,1}(X,Y,Z) \,\mathrm{d}Y\mathrm{d}X=0.\label{BC ND FO De3}
\end{equation}
\end{subequations}It is worth noting that, as Eq.~(\ref{Momentum FO De})
indicates, the determination of the pressure distribution at the first
order in $De$ only requires  calculating $\mathcal{T}_{zz,1}$ and $\mathcal{T}_{yz,1}$,
but not $\mathcal{T}_{yy,1}$. 

Substituting Eqs.~(\ref{Tzz FO De})--(\ref{Tyz FO De}) into Eq.~(\ref{Momentum FO De}), we obtain
\begin{equation}
\frac{\mathrm{d}P_{1}}{\mathrm{d}Z}-\dfrac{\partial^{2}V_{z,1}}{\partial Y^{2}}=\tilde{\beta}S(X,Y,Z),\label{Momentum FO De e}
\end{equation}
where we have defined 
 \begin{multline}
\tilde{\beta}S(X,Y,Z)\triangleq\frac{\partial}{\partial Z}\left(2\dfrac{\partial V_{z,0}}{\partial Y}\mathcal{T}_{yz,0}\right)\\
-\frac{\partial}{\partial Y}\left(V_{y,0}\frac{\partial\mathcal{T}_{yz,0}}{\partial Y}+V_{z,0}\frac{\partial\mathcal{T}_{yz,0}}{\partial Z}-\dfrac{\partial V_{z,0}}{\partial Y}\mathcal{T}_{yy,0}\right).\label{beta*S(X,Y,Z) FO De}
\end{multline}
Note that since $\partial\mathcal{T}_{yz,0}/\partial Y=-12\tilde{\beta}/H_{e}(Z)^{3}$ is solely a function of $Z$, the first term appearing in the second row of Eq.~(\ref{beta*S(X,Y,Z) FO De}) can be expressed as
\begin{equation}
\frac{\partial}{\partial Y}\left(V_{y,0}\frac{\partial\mathcal{T}_{yz,0}}{\partial Y}\right)=\frac{\partial V_{y,0}}{\partial Y}\frac{\partial\mathcal{T}_{yz,0}}{\partial Y}=-\frac{\partial V_{z,0}}{\partial Z}\frac{\partial\mathcal{T}_{yz,0}}{\partial Y}.\label{FT FO De e}
\end{equation}
Thus, from Eqs.~(\ref{Momentum FO De e})--(\ref{FT FO De e}), it follows that finding the pressure distribution at the first order in $De$ does not require determining the explicit expression for $V_{y,0}$.

The right-hand side of Eq.~(\ref{Momentum FO De e}) only depends on the leading-order solution, and thus can be explicitly calculated using Eqs.~(\ref{Axial vel LO De e}), (\ref{Tyy  LO e}), (\ref{Tyz LO e}),
and (\ref{FT FO De e}) to yield,
\begin{equation}
S(X,Z)=\frac{36H(X,Z)}{H_{e}(Z)^{6}}\frac{\partial H}{\partial Z}-\frac{108H_{e}'(Z)H(X,Z)^{2}}{H_{e}(Z)^{7}}.\label{S(X,Z) FO De e}
\end{equation}
Similar to the two-dimensional pressure-driven flow of an Oldroyd-B fluid in a rigid non-uniform channel \citep{boyko2022pressure}, the source term $S(X,Z)$  given in Eq.~(\ref{S(X,Z) FO De e}), arising at the first order in $De$, is independent of the transverse coordinate $Y$.

Next, integrating Eq.~(\ref{Momentum FO De e}) twice with respect to $Y$, using Eq.~(\ref{S(X,Z) FO De e}), and applying the no-slip boundary conditions from Eq.~(\ref{BC ND FO De1}), we obtain
\begin{equation}
V_{z,1}(X,Y,Z)=-\frac{1}{2}\left[\frac{\mathrm{d}P_{1}}{\mathrm{d}Z}-\tilde{\beta}S(X,Z)\right]Y(H-Y).\label{Axial vel FO De}
\end{equation}
To determine $\mathrm{d}P_{1}/\mathrm{d}Z$, we use the integral constraint from Eq.~(\ref{BC ND FO De3}) to find
\begin{gather}
 \int_{-1/2}^{+1/2}\left[\frac{\mathrm{d}P_{1}}{\mathrm{d}Z}-\tilde{\beta}S(X,Z)\right]H(X,Z)^{3} \,\mathrm{d}X=0 \nonumber\\ 
  \Rightarrow\frac{\mathrm{d}P_{1}}{\mathrm{d}Z}=\tilde{\beta}\frac{\mathcal{S}(Z)}{H_{e}(Z)^{3}},\label{dP/dZ FO De e}
\end{gather}
where we have defined
\begin{equation}
\begin{aligned}
\mathcal{S}(Z) &\triangleq\int_{-1/2}^{+1/2}S(X,Z)H(X,Z)^{3} \,\mathrm{d}X \\ 
  &=\frac{36}{H_{e}(Z)^{6}}\int_{-1/2}^{+1/2}H(X,Z)^{4}\frac{\partial H}{\partial Z} \,\mathrm{d}X\\
  &\phantom{=}~ -\frac{108H_{e}'(Z)}{H_{e}(Z)^{7}}\int_{-1/2}^{+1/2}H(X,Z)^{5} \,\mathrm{d}X.\label{S(Z) FO De e}
\end{aligned}
\end{equation}
Observe that, for a given shape $H$ of the top wall of the channel, Eq.~(\ref{dP/dZ FO De e}) provides an explicit expression for the pressure gradient correction at the first order in $De$.

\subsection{Summary}

In Secs. \ref{LO section} and \ref{FO section}, we derived
dimensionless closed-form expressions for the axial  pressure gradient as a function
of the channel's shape function $H(X,Z)$ (and $H_{e}(Z)$), the viscosity ratio
$\tilde{\beta}$, and the Deborah number $De$ up to $\mathrm{O}(De)$. Specifically,
\begin{equation}
\frac{\mathrm{d}P_{0}}{\mathrm{d}Z}=-\frac{12}{H_{e}(Z)^{3}},\label{dP/dZ LOe}
\end{equation}
\begin{equation}
\frac{\mathrm{d}P_{1}}{\mathrm{d}Z}=\tilde{\beta}\frac{\mathcal{S}(Z)}{H_{e}(Z)^{3}},\label{dP/dZ FOe}
\end{equation}
where the expressions for $H_{e}(Z)$ and $\mathcal{S}(Z)$ are given
in Eqs.~(\ref{He(Z)}) and (\ref{S(Z) FO De e}), respectively.

It is worth noting that Eqs.~(\ref{dP/dZ LOe}) and (\ref{dP/dZ FOe}) are not restricted to deformable channels. Indeed, for a three-dimensional non-uniform rigid channel, whose top wall has a prescribed shape $H(X,Z)$, Eqs.~(\ref{dP/dZ LOe}) and (\ref{dP/dZ FOe}) can be combined as
\begin{equation}
\frac{\mathrm{d}P}{\mathrm{d}Z}=-\frac{1}{H_{e}(Z)^{3}}\left[12-De\tilde{\beta}\mathcal{S}(Z)\right]+\mathrm{O}(De^{2}),\label{dP/dZ LO and FO De e3}
\end{equation} 
thus providing the leading-order effect of viscoelasticity on the pressure drop.

\section{Coupling hydrodynamics to elasticity}\label{Coupling}

In this section, we couple Eqs.~(\ref{dP/dZ LOe})--(\ref{dP/dZ FOe}), obtained from the hydrodynamic problem, to the elastic deformation of the top wall of the channel, and then we solve the resulting elastohydrodynamic problem for the Oldroyd-B fluid in the weakly viscoelastic limit.

Using the non-dimensionalization given in Eq.~(\ref{ND variables}), the dimensionless deformed height of the fluid domain, $H(X,Z)$, can be expressed in terms of the dimensionless top wall deformation, $U(X,Z)$, as $H(X,Z)=1+\alpha U(X,Z)$,
where $\alpha\triangleq u_{c}/h_{0}$ is the dimensionless number that quantifies the compliance of the top wall.

Recent studies have shown that the deformations of the $(X, Y)$ cross-section of the channel at different $Z$-locations decouple from each other, leading to a local deformation--pressure relation \cite{christov2018flow,shidhore2018static,wang2019theory,wang2020soft,wang2021reduced}. At each cross-section, the deformation is determined by the local pressure
$P(Z)$, and therefore, in general,  we can express the dimensionless deformed shape of the channel as
\begin{equation}
H(X,Z)=1+\alpha U(X,Z)=1+\alpha F(X)P(Z).\label{H(X,Z)}
\end{equation}
Here, the spanwise profile $F(X)$ is obtained by solving the corresponding
elasticity problem in an  $(X,Y)$ cross-section, with suitable boundary conditions \cite{christov2018flow,shidhore2018static,wang2019theory}. In Table~\ref{T2}, we summarize the known expressions for the spanwise profiles $F(X)$ and list explicitly the formulas for $\alpha$ for three different (representative) elastic models of the top wall deformation.

\begin{table*}[width=.9\textwidth,cols=4,pos=h]
    \caption{Summary of expressions for the dimensionless shape function, $F(X)=U(X,Z)/P(Z)$, from which the top wall deformation is obtained, and the corresponding compliance parameter, $\alpha=u_{c}/h_{0}$, for three representative elasticity models from the literature.}
    \centering
    \begin{tabular*}{\tblwidth}{@{} LLLL@{} }
    \toprule
    Elastic model&  Dimensionless top wall deformation & Compliance parameter   & Reference\\
 & shape in cross-section $F(X)$ & $\alpha=u_{c}/h_{0}$  & \\
\midrule
\midrule

Kirchhoff--Love plate-bending theory   &   $\displaystyle{\left(\frac{1}{4}-X^{2}\right)^{2}}$  &  $ \displaystyle{\frac{(1-\nu^{2})\eta_0 qw^{3}\ell}{2E_{Y} b^{3} h_{0}^{4}}}$  & \cite{christov2018flow}\\
($u_{\mathrm{max}}\ll b \ll w$, $b/w \to 0$)  &      &    & \\

 
\midrule

 Reissner--Mindlin plate-bending theory    &   $ \displaystyle{\left(\frac{1}{4}-X^{2}\right)\left[\frac{2(b/w)^{2}}{\kappa(1-\nu)}+\left(\frac{1}{4}-X^{2}\right)\right]}$    &  $ \displaystyle{\frac{(1-\nu^{2})\eta_0 qw^{3}\ell}{2E_{Y} b^{3} h_{0}^{4}}}$  & \cite{shidhore2018static,wang2021reduced}\\
 
($u_{\mathrm{max}}\ll b < w$, $b/w\not\to0$)   &       &   & \\

\midrule

Large-thickness linear elasticity theory & $\displaystyle{\frac{2}{\pi}\sum_{m=1}^{\infty}\frac{A_{m}}{m}\sin\left[m\pi\left(X+\frac{1}{2}\right)\right]},$   &            $\displaystyle{\frac{(1-\nu^{2})\eta_0 q \ell}{E_{Y}h_{0}^4}}$ & \cite{wang2019theory,wang2021reduced} \\
($u_{\mathrm{max}}\ll w\ll b$, $b/w\to\infty$) & $\displaystyle{A_{m}=\frac{2}{m\pi}\left[1-(-1)^{m}\right]}$   &             & \\
    \bottomrule
    \end{tabular*}
    \label{T2}
\end{table*}

Before coupling Eqs.~(\ref{dP/dZ LOe})--(\ref{dP/dZ FOe}) and Eq.~(\ref{H(X,Z)}), it is worth noting that the perturbative form of the solution for the pressure, as $P(Z)=P_{0}(Z)+De P_{1}(Z)+\mathrm{O}(De^2)$, necessarily leads to a perturbation expansion of the channel shape $H(X, Z)$ in powers of $De$ as well. Substituting the latter expansion for the pressure into Eq.~(\ref{H(X,Z)}) yields
\begin{equation}
H(X,Z)=\underbrace{1+\alpha F(X)P_{0}(Z)}_{H_{0}(X,Z)}
+De\underbrace{\alpha F(X) P_{1}(Z)}_{H_{1}(X,Z)}.\label{H(X,Z) expand}
\end{equation}
Next, substituting Eq.~(\ref{H(X,Z) expand}) into Eqs.~(\ref{dP/dZ LOe})--(\ref{dP/dZ FOe}) and using Eqs.~(\ref{He(Z)}) and (\ref{S(Z) FO De e}), we obtain
\begin{equation}
\frac{\mathrm{d}P_{0}}{\mathrm{d}Z}=-\frac{12}{H_{e,0}(Z)^{3}},\label{dP/dZ LO expansion}
\end{equation}
\begin{equation}
\frac{\mathrm{d}P_{1}}{\mathrm{d}Z}=\tilde{\beta}\frac{\mathcal{S}_{0}(Z)}{H_{e,0}(Z)^{3}},\label{dP/dZ FO expansion}
\end{equation}
where $\mathcal{S}_{0}(Z)$ and $H_{e,0}(Z)$ are functions of $P_{0}(Z)$
and $\alpha$, given by
\begin{multline}
\mathcal{S}_{0}(Z)=\frac{\mathrm{d}P_{0}}{\mathrm{d}Z}\frac{36\alpha}{H_{e,0}(Z)^{6}}\left[\mathcal{I}_{1}+4\alpha\mathcal{I}_{2}P_{0}(Z) \right.\\
 \left. +6\alpha^{2}\mathcal{I}_{3}P_{0}(Z)^{2}+4\alpha^{3}\mathcal{I}_{4}P_{0}(Z)^{3}+\alpha^{4}\mathcal{I}_{5}P_{0}(Z)^{4}\right]\\
    -\frac{108H_{e,0}'(Z)}{H_{e,0}(Z)^{7}}\left[ 1+5\alpha\mathcal{I}_{1}P_{0}(Z)+10\alpha^{2}\mathcal{I}_{2}P_{0}(Z)^{2}\right.\\
   \left. +10\alpha^{3}\mathcal{I}_{3}P_{0}(Z)^{3}+5\alpha^{4}\mathcal{I}_{4}P_{0}(Z)^{4}+\alpha^{5}\mathcal{I}_{5}P_{0}(Z)^{5}\right] \label{S(Z)0}
\end{multline} 
and
\begin{multline}
H_{e,0}(Z)=\left[1+3\alpha\mathcal{I}_{1}P_{0}(Z) \right.\\
 \left. +3\alpha^{2}\mathcal{I}_{2}P_{0}(Z)^{2}+\alpha^{3}\mathcal{I}_{3}P_{0}(Z)^{3}\right]^{1/3},\label{He,0}
\end{multline}
where the constants $\mathcal{I}_{i}$ are defined as 
\begin{equation}
\mathcal{I}_{i}=\int_{-1/2}^{+1/2}F(X)^{i} \,\mathrm{d}X,\qquad i=1,2,3,\hdots\,.\label{I coeff.}
\end{equation}
We note that the right-hand side of Eq.~(\ref{dP/dZ FO expansion}) for the first-order pressure gradient depends only on the leading-order Newtonian solution for the pressure and deformation, consistent with a regular asymptotic expansion in $De$.

At the leading order in $De$, substituting Eq.~(\ref{He,0}) into Eq.~(\ref{dP/dZ LO expansion}), we obtain a separable first-order ordinary differential equation for $P_{0}(Z)$. The implicit solution for $P_{0}(Z)$ (see, e.g., \cite{wang2021reduced}), subject to $P_{0}(1)=0$, is 
\begin{multline}
12(1-Z)=P_{0}(Z)\left[1+\frac{3}{2}\alpha\mathcal{I}_{1}P_{0}(Z) \right.\\
 \left. +\alpha^{2}\mathcal{I}_{2}P_{0}(Z)^{2}+\frac{1}{4}\alpha^{3}\mathcal{I}_{3}P_{0}(Z)^{3}\right].\label{P0(Z) imp solution}
\end{multline}
Unlike for the leading-order problem, it is difficult to obtain an analytical closed-form solution for the first-order correction to the pressure distribution, $P_{1}(Z)$, since, generally, the right-hand side of Eq.~(\ref{dP/dZ FO expansion}) is a complicated function of $Z$. Therefore, we numerically solved Eqs.~(\ref{dP/dZ LO expansion})--(\ref{dP/dZ FO expansion}) using \textsc{Matlab}'s routine \texttt{ode45}.

\begin{figure*}[t]
    \centering
    \includegraphics[width=0.8\textwidth]{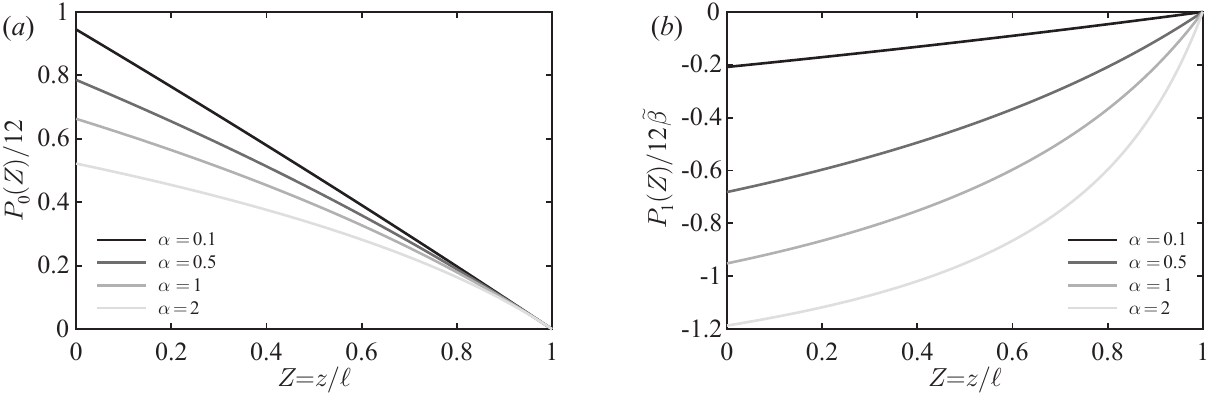}
    \caption{ (\textit{a}) The leading- and (\textit{b}) first-order-in-$De$ contributions to the pressure distribution, $P_{0}(Z)/12$ and $P_{1}(Z)/(12\tilde{\beta})$, respectively, as functions of the axial coordinate $Z=z/\ell$, for an Oldroyd-B fluid in a channel with a deformable top wall, modeled using the plate-bending theory, for different values of the compliance parameter $\alpha$.}
    \label{F2}
\end{figure*}

To provide further analytical insight, we consider the case of a weakly
deformable channel and expand the right-hand
side of Eqs.~(\ref{dP/dZ LO expansion}) and (\ref{dP/dZ FO expansion}) into a Taylor series for $\alpha\ll1$, while also using Eqs.~(\ref{S(Z)0})--(\ref{He,0}),
to obtain\begin{subequations}
\begin{align}
\frac{\mathrm{d}P_{0}}{\mathrm{d}Z}&=-12[1-3\alpha\mathcal{I}_{1}P_{0}(Z)]+\mathrm{O}(\alpha^{2}),\label{dP/dZ LO expansion small alpha}\\
\frac{\mathrm{d}P_{1}}{\mathrm{d}Z}&=-72\tilde{\beta}\alpha\mathcal{I}_{1}\frac{\mathrm{d}P_{0}}{\mathrm{d}Z}+\mathrm{O}(\alpha^{2}).\label{dP/dZ FO expansion small alpha}
\end{align}
\end{subequations}The solution of Eq.~(\ref{dP/dZ LO expansion small alpha}),
subject to $P_{0}(1)=0$, is 
\begin{equation}
P_{0}(Z)=\frac{1-\exp\left[-36\alpha\mathcal{I}_{1}(1-Z)\right]}{3\alpha\mathcal{I}_{1}}.\label{dP/dZ LO expansion small alpha sol}
\end{equation}
Substituting Eq.~(\ref{dP/dZ LO expansion small alpha sol}) into
Eq.~(\ref{dP/dZ FO expansion small alpha}) and solving subject to $P_{1}(1)=0$ yields 
\begin{equation}
P_{1}(Z)=-24\tilde{\beta}\left\{1-\exp\left[-36\alpha\mathcal{I}_{1}(1-Z)\right]\right\}.\label{dP/dZ FO expansion small alpha sol}
\end{equation}
Accordingly, the dimensionless pressure drop, up to $\mathrm{O}(\alpha De)$, is obtained by taking $Z=0$ in Eqs.~(\ref{dP/dZ LO expansion small alpha sol})
and (\ref{dP/dZ FO expansion small alpha sol}):
\begin{align}\label{Pressure drop  up to O(alpah*De)}
\Delta P &= \left[1-\exp\left(-36\alpha\mathcal{I}_{1}\right)\right]\left[\frac{1}{3\alpha\mathcal{I}_{1}}-24\tilde{\beta}De\right]\\
 &\approx12[1-18\mathcal{I}_{1}(\alpha+4\tilde{\beta}\alpha De)]+\mathrm{O}(\alpha^{2},\alpha^{2}De,\alpha De^{2}).\nonumber
\end{align}
Equation~(\ref{Pressure drop  up to O(alpah*De)}) clearly shows that the leading-order combined effect of the compliance of the elastic channel (with $\alpha>0$ for inflated channels, as in the present work) and the viscoelasticity of the fluid, is to decrease the dimensionless pressure drop.

Finally, we note that in Eq.~(\ref{Pressure drop  up to O(alpah*De)}) there is no $\mathrm{O}(De)$ term (only the mixed $\mathrm{O}(\alpha De)$ term) because the leading-order viscoelastic correction to the pressure drop of the Oldroyd-B fluid in a straight rigid channel vanishes. 
Specifically, this absence of the $\mathrm{O}(De)$ term is because, in a \emph{rigid} uniform channel, the flow and pressure fields of the Oldroyd-B fluid are the same as those of a Newtonian fluid with a constant viscosity $\eta_{0}$.

\section{Results and discussion}\label{Results}

In this section, we present the results illustrating the leading-order effect of viscoelasticity on the pressure drop and on the deformed channel shape due to the flow of an Oldroyd-B fluid in a slender and shallow compliant channel.

As an illustrative example, we model the compliance of the top wall using the plate-bending theory (the first elastic model in Table~\ref{T2}), which holds under the assumptions that the maximum displacement of the top wall $u_\mathrm{max}$ is small compared to its thickness $b$, and the thickness $b$ is small compared to its width $w$, i.e., $u_\mathrm{max}\ll b\ll w$ \cite{timoshenkoPlates,howell2009applied}. The corresponding values of $\{\mathcal{I}_{1},\hdots,\mathcal{I}_{5}\}$ for the plate-bending theory are summarized in Table~\ref{T3}. We note that the other two elastic models from Table \ref{T2} yield qualitatively similar results since the difference lies only in the values of $\{\mathcal{I}_{1},\hdots,\mathcal{I}_{5}\}$. 

\begin{table}[width=\linewidth,cols=4,pos=b]
    \caption{Values of the coefficients $\{\mathcal{I}_{1},\hdots,\mathcal{I}_{5}\}$
defined by Eq.~(\ref{I coeff.}) for a rectangular channel with a deformable top wall modeled using  the Kirchhoff--Love plate-bending theory.}
    \begin{tabular*}{\tblwidth}{@{} CCCCC@{} }
    \toprule    
   $\mathcal{I}_{1}$ &  $\mathcal{I}_{2}$ &  $\mathcal{I}_{3}$ &  $\mathcal{I}_{4}$ &  $\mathcal{I}_{5}$ \\
      \midrule
    \midrule
    $\dfrac{1}{30}$ & $\dfrac{1}{630}$ & $\dfrac{1}{12012}$ & $\dfrac{1}{218790}$& $\dfrac{1}{3879876}$\\
    \bottomrule
    \end{tabular*}
    \label{T3}
\end{table}
\begin{figure}[th]
    \centering
    \includegraphics[width=0.68\linewidth]{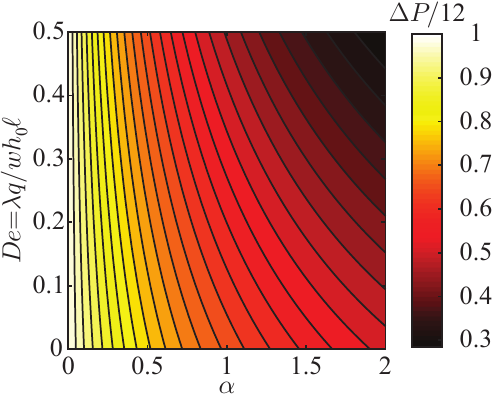}
    \caption{Contour plot of the dimensionless pressure drop $\Delta P/12=\Delta pwh_{0}^{3}/12\eta_0 q\ell$ for the Oldroyd-B fluid in a  channel with a deformable top wall, modeled using  the Kirchhoff--Love plate-bending theory, as a function of $\alpha$ and $De$ with $\tilde{\beta}=0.4$.}
    \label{F3}
\end{figure}

To illustrate the representative values of the physical parameters in the flow-rate-controlled flow of an Oldroyd-B
fluid in a deformable channel, we consider an experimental configuration studied by Ozsun et al.~\cite{OYE13}. Specifically, the rectangular microchannel has a length $\ell=15.5$ mm, width $w=1.7$ mm, and undeformed height $h_0=0.244$ mm, whose top wall has thickness $b=0.2$ mm, Young's modulus $E_Y\approx 1.6$ MPa, and Poisson's ratio $\nu=0.499$. 
In particular, we observe that $\epsilon=0.016$ and $\delta=0.144$, which satisfy $\epsilon\ll\delta\ll1$, and $b\ll w$, allowing the use of the Kirchhoff--Love plate-bending theory. As the
working fluid, consider a $0.01\%$ by weight solution of a high molecular weight polyacrylamide (PAA), used by Groisman \& Quake~\cite{groisman2004microfluidic}. The corresponding relaxation time is $\lambda \approx 13$~ms, and the solution and solvent viscosities are $\eta_0=1.92$ mPa s and $\eta_s=1.37$ mPa s, respectively.

For flow rates in the range of $q=1-50$ mL $\mathrm{min}^{-1}$, which are feasible in experiments (see, e.g., \cite{OYE13}), we obtain that $\alpha=0.02-1$, $De=0.034-1.69$, and $Wi=2.14-107$. Therefore, the low-Deborah-lubrication analysis considered in this work is relevant to real situations and can be realized in experiments. Note that for $De\ll1$ we have $Wi =\mathrm{O}(1)$.

Next, in Figs.~\ref{F2}(\textit{a},\textit{b}), we present the leading- and first-order contributions in $De$ to the pressure distribution for the Oldroyd-B fluid in a deformable channel, i.e., $P_{0}(Z)/12$ and $P_{1}(Z)/(12\tilde{\beta})$ obtained from Eqs.~(\ref{dP/dZ LO expansion}) and (\ref{dP/dZ FO expansion}), respectively, as functions of $Z$ for different values of $\alpha$. We choose to present $P_{1}(Z)/(12\tilde{\beta})$ rather than $P_{1}(Z)/12$, since $P_{1}(Z)/(12\tilde{\beta})=\mathcal{S}_{0}(Z)/(12H_{e,0}(Z)^{3})$ is independent of $\tilde{\beta}$, so that $P_{0}(Z)/12$ and $P_{1}(Z)/(12\tilde{\beta})$ are general expressions for the Oldroyd-B fluid, which depend only on $\alpha$ and the elastic model for the top wall (via the values of $\{\mathcal{I}_{1},\hdots,\mathcal{I}_{5}\}$).

It is evident from Fig.~\ref{F2}(\textit{a}) that $P_{0}(Z)$, corresponding to the Newtonian pressure distribution, is positive and shows a sub-linear behavior, which is more pronounced as $\alpha$ increases. On the other hand, Fig.~\ref{F2}(\textit{b}) illustrates that $P_{1}(Z)/(12\tilde{\beta})$, corresponding to the leading-order effect of the viscoelasticity on the pressure, is negative and shows a super-linear behavior, further decreasing as $\alpha$ increases. Thus, both of these contributions lead to a reduction in the dimensionless pressure drop, $\Delta P=P_{0}(0)+De P_{1}(0)$, with increasing $\alpha$, for given values of $De$ and $\tilde{\beta}$, consistent with the analytical prediction in Eq.~(\ref{Pressure drop  up to O(alpah*De)}).
\begin{figure}[ht]
    \centering
    \includegraphics[width=0.8\linewidth]{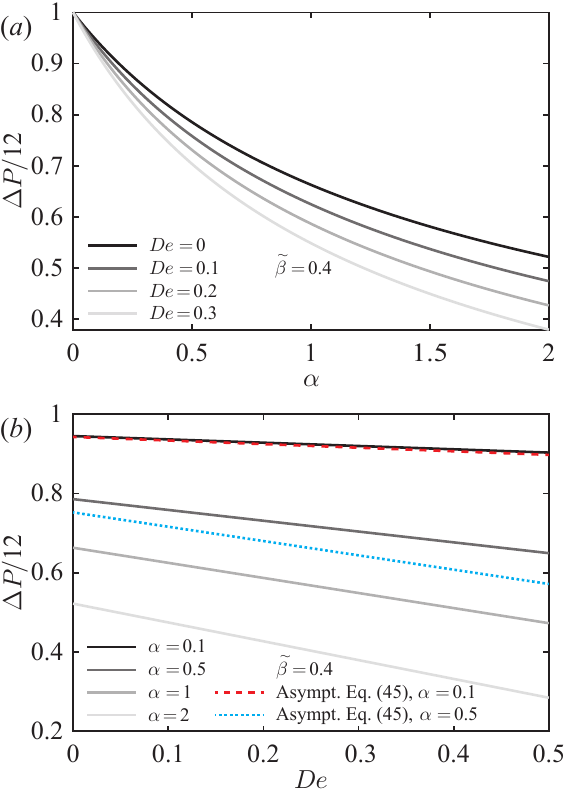}
    \caption{The dimensionless pressure drop $\Delta P/12=\Delta pwh_{0}^{3}/12\eta_0 q\ell$ for the flow of an Oldroyd-B fluid in a channel with a deformable top wall, modeled using the plate-bending theory. (\textit{a}) $\Delta P/12$ as a function of $\alpha$ for different values of $De$ with $\tilde{\beta}=0.4$. (\textit{b}) $\Delta P/12$ as a function of $De$ for different values of $\alpha$ with $\tilde{\beta}=0.4$. Solid curves represent the dimensionless pressure drop, obtained from numerically solving Eqs.~(\ref{dP/dZ LO expansion})--(\ref{dP/dZ FO expansion}). Dashed and dotted curves in (\textit{b}) represent the asymptotic solution given in Eq.~(\ref{Pressure drop  up to O(alpah*De)}) for $\alpha=0.1$ and $\alpha=0.5$, respectively.}
    \label{F4}
\end{figure}

In Fig.~\ref{F3}, we present a contour plot of the dimensionless pressure drop $\Delta P/12=\Delta p w h_{0}^{3}/12\eta_0 q\ell$ for the flow of an Oldroyd-B fluid in the deformable channel with an elastic plate-like top wall, as a function of $\alpha$ and $De$ for $\tilde{\beta}=0.4$, obtained from numerically solving Eqs.~(\ref{dP/dZ LO expansion})--(\ref{dP/dZ FO expansion}). Fig.~\ref{F3} clearly indicates that both compliance of the deforming top wall and viscoelasticity of fluid decrease the dimensionless pressure drop, consistent with the analytical prediction in Eq.~(\ref{Pressure drop  up to O(alpah*De)}), which is strictly valid for $\alpha\ll1$ and $De\ll1$.

For further clarification, Fig.~\ref{F4}(\textit{a}) presents the dimensionless pressure drop $\Delta P/12=\Delta pwh_{0}^{3}/12\eta_0 q\ell$ as a function of $\alpha$, for different values of $De$, and Fig.~\ref{F4}(\textit{b}) presents the dimensionless pressure drop~$\Delta P/12=\Delta pwh_{0}^{3}/12\eta_0 q\ell$ as a function of $De$ for different values of $\alpha$. Solid curves represent the dimensionless pressure drop accounting for the first-order-in-$De$ correction due to viscoelasticity, obtained from numerically solving Eqs.~(\ref{dP/dZ LO expansion})--(\ref{dP/dZ FO expansion}). Dashed and dotted curves in Fig.~\ref{F4}(\textit{b}) represent the asymptotic solutions for $\alpha=0.1$ and $\alpha=0.5$, respectively, given by Eq.~(\ref{Pressure drop  up to O(alpah*De)}).


It follows from Fig.~\ref{F4}(\textit{a}) that the dimensionless pressure drop decreases nonlinearly with $\alpha$, with the behavior becoming more pronounced as $De$ increases, consistent with the results shown in Fig.~\ref{F3}. Unlike the \emph{nonlinear} dependence on $\alpha$, our first-order-in-$De$ theory predicts that $\Delta P/12$ decreases \emph{linearly} with $De$ for $De\ll1$, as shown in Fig.~\ref{F4}(\textit{b}). Further investigation that involves the numerical solution of a set of lubrication equations (\ref{LO Lub Eq}) coupled to the elasticity through Eq.~(\ref{H(X,Z)}) is required to assess the behavior of the pressure drop at higher values of $De$ and is beyond the scope of this work. It is evident from Fig.~\ref{F4}(\textit{b}) that the asymptotic solution Eq.~(\ref{Pressure drop  up to O(alpah*De)}), which is strictly valid for $\alpha\ll1$ and $De\ll1$, slightly underpredicts the numerical solution. Yet, even for $\alpha=0.5$, it results in a modest relative error of approximately  $12\%$ at $De=0.5$.

\begin{figure*}[h]
    \centering
    \includegraphics[width=0.8\textwidth]{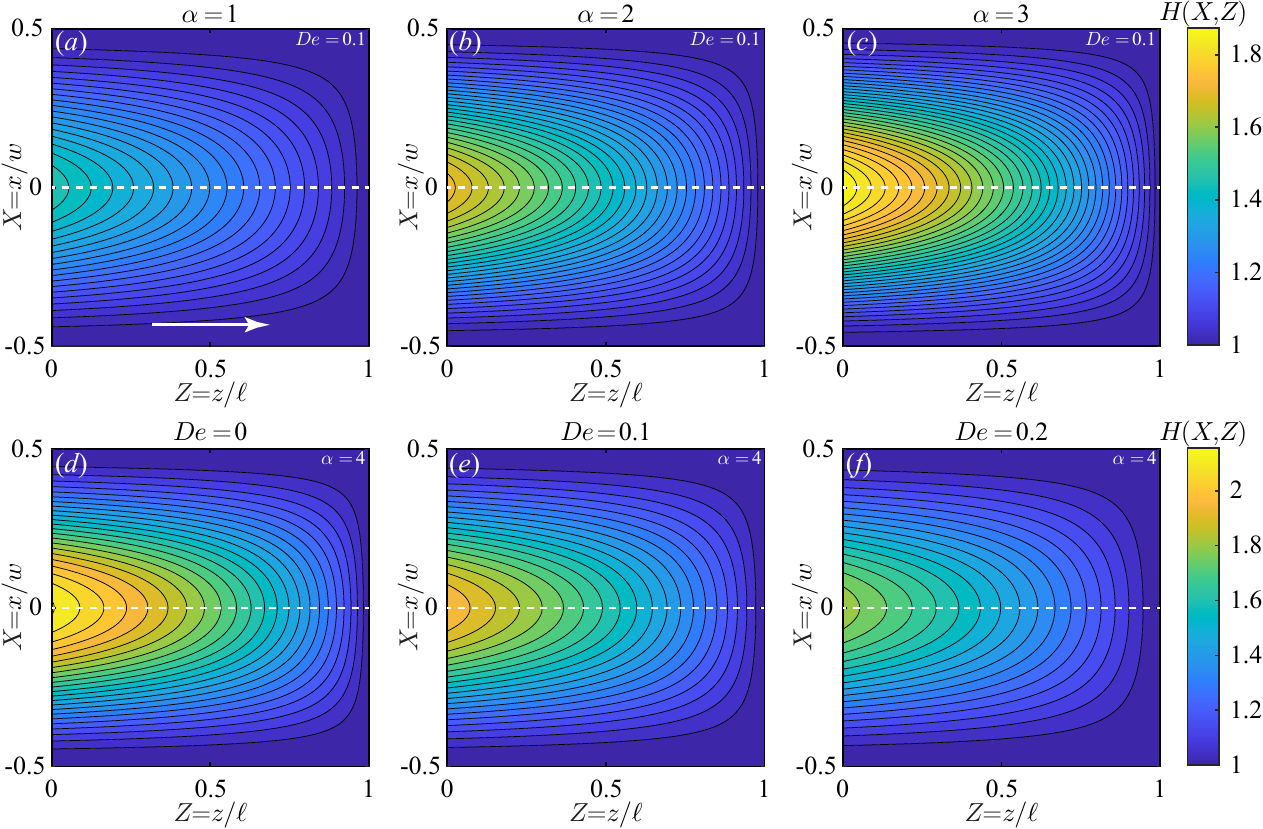}
    \caption{The dimensionless deformed height, $H(X,Z)$, of the channel shown as a contour plot in the $(X, Z)$ plane for (\textit{a})--(\textit{c}) $\alpha=1,2,3$, with $De=0.1$ and $\tilde{\beta}=0.4$, and for (\textit{d})--(\textit{f}) $De=0,0.1,0.2$, with $\alpha=4$ and $\tilde{\beta}=0.4$. White dashed lines represent the centerline, $X=0$, where the deformation is maximal. Arrow indicates the flow direction.}
    \label{F5}
\end{figure*}

In addition to the pressure drop, it is of  interest to understand the effect of viscoelasticity on the deformation of the channel's top wall. To this end, in Figs.~\ref{F5}(\textit{a}--\textit{c}), we show contours of the dimensionless deformed shape of the channel, $H(X,Z)$, for different values of $\alpha$, with $De=0.1$ and $\tilde{\beta}=0.4$. In Figs.~\ref{F5}(\textit{d}--\textit{f}), we illustrate contours of the dimensionless deformed shape of the channel, $H(X,Z)$, as a function of the $(X,Z)$ coordinates for different values of $De$, for $\alpha=4$ and $\tilde{\beta}=0.4$. We clearly observe the existence of the trade-off between the effect of compliance of the channel's wall and the effect of viscoelasticity of the fluid on the maximum displacement of the deforming top wall. While, as expected, the maximum displacement, i.e., $H(0,0)=1+\alpha F(0)P(0)$, increases with $\alpha$, it decreases with $De$ due to the reduction in the pressure $P(0)=\Delta P$, consistent with the results shown in Figs.~\ref{F3} and \ref{F4}.

Finally, Fig.~\ref{F5} indicates that the deformation $H(0, Z)$ (white dashed curves) monotonically decreases as a function of $Z$, meaning that the Oldroyd-B fluid flows at a steady state through a contracting (deformed) channel. Thus, the predicted reduction in the dimensionless pressure drop at low $De$, associated with viscoelasticity, is consistent with previous numerical and analytical results for Oldroyd-B fluids flowing through rigid contracting channels \cite{binding2006contraction,varchanis2021reduced,boyko2022pressure}.

\section{Concluding remarks}\label{CR}

In this work, we analyzed the pressure-driven flow of a weakly viscoelastic Oldroyd-B fluid in slender and shallow deformable channels and developed a theoretical framework for calculating the pressure, deformation and the flow rate--pressure drop relation due to the fluid--structure interaction. We first identified the appropriate characteristic scales and dimensionless parameters governing a viscoelastic flow in a slender and shallow geometry under the lubrication approximation. Then, employing a perturbation expansion in powers of the Deborah number $De$ and coupling the hydrodynamic problem to the elastic problem via the deformation of the top wall, we  solved the elastohydrodynamic problem. Specifically, we obtained closed-form expressions for the pressure gradient and deformation, leading to the $q$ -- $\Delta p$ relation in the weakly viscoelastic limit, up to $\mathrm{O}(De)$, and valid for arbitrary values of the compliance parameter $\alpha$.

For an elastic top wall, whose deformation obeys the Kirchhoff--Love plate-bending theory, our key analytical result, Eq.~(\ref{Pressure drop  up to O(alpah*De)}), for the flow rate--pressure drop relation in dimensional form, accounting for the leading-order effects of viscoelasticity of the fluid and compliance of the elastic wall, $De\ll1$ and $\alpha\ll1$, is:
\begin{multline}
\Delta p\approx\frac{12\eta_{0}q\ell}{wh_{0}^{3}}\\
 \times\left[1-\frac{3}{10}\frac{(1-\nu^{2})\eta_{0}q\ell w^{3}}{E_{Y}h_{0}^{4}b^{3}}\left(1+4\frac{\eta_{p}}{\eta_{0}}\frac{\lambda q}{wh_{0}\ell}\right)\right],\label{q-dp dim bending}
\end{multline}
where $q$ is the flow rate, $\Delta p$ is the pressure drop, $w$ is the channel width, $h_0$ is the
undeformed channel height, $\ell$ is the channel length, $b$ is the top wall's thickness, $E_{Y}$ and $\nu$ are, respectively, the Young's modulus and Poisson's ratio of the elastic material,  $\eta_{p}$ is the polymer contribution to the zero-shear-rate viscosity,
$\eta_{0}$ is the total zero-shear-rate viscosity, and $\lambda$ is the polymer relaxation time. Equation~(\ref{q-dp dim bending}) clearly shows that the leading-order combined effect of the compliance of the deforming top wall and the viscoelasticity of the fluid is to decrease the pressure drop for a given flow rate. We note that similar expressions to Eq.~(\ref{q-dp dim bending}) for the flow rate--pressure drop relation can be obtained from Eq.~(\ref{Pressure drop  up to O(alpah*De)}) for the other two wall deformation models using the expressions for $\alpha$ and $F(X)$ given in Table~\ref{T2}.

Our theoretical approach is not limited to the case of a three-dimensional channel of an initially rectangular cross-section, and it can also be used to study the pressure-driven flow of viscoelastic fluids in slender axisymmetric deformable tubes. As the fluid--structure interaction between these geometries and the shear-thinning fluids has previously received considerable attention (see Table~\ref{T1}), it would be interesting to understand how viscoelasticity influences the interplay between the complex fluid flow and the compliance of the elastic tube at low Reynolds number.

Another interesting extension of the present work, which has assumed vanishing fluid inertia, is to include weak but finite inertia, $\hat{Re}=\epsilon Re=\mathrm{O}(1)$, where $\hat{Re}=\epsilon \rho v_{c} h_{0}/\eta_{0}$ is the reduced Reynolds number and $\rho$ is the fluid's density. Recent studies showed that accounting for the finite inertia of  a Newtonian fluid leads to a pressure drop increase in a deformable microchannel \cite{wang2020soft,wang2021reduced}. Noteworthy, the resulting expression for the pressure drop for the Newtonian fluid incorporating the flow's inertia involved all the $\{\mathcal{I}_{1},\hdots,\mathcal{I}_{5}\}$ coefficients \cite{wang2020soft}, similar to Eqs.~(\ref{dP/dZ LO expansion})--(\ref{dP/dZ FO expansion}) for the dimensionless pressure drop of the inertialess flow of an Oldroyd-B fluid. Therefore, in future work, it would be of particular interest to explore the combined influence of fluid viscoelasticity and fluid inertia, as well as their interplay, on the pressure drop in a deformable channel.

\section*{Acknowledgement}
E.B.\ acknowledges the support of the Lillian Gilbreth Postdoctoral Fellowship from Purdue's College of Engineering and the Zuckerman STEM Leadership Program. I.C.C.\ acknowledges partial support by the US National Science Foundation (NSF) under grant No.\ CBET-1705637.



\bibliographystyle{elsarticle-num}

\balance


\end{document}